\documentclass[5p,times]{elsarticle}
\usepackage{lipsum}
\usepackage{amsmath,amssymb,amsfonts}
\usepackage{algorithmic}
\usepackage{textcomp}
\usepackage{xcolor}
\usepackage[T1]{fontenc}
\usepackage{graphicx}
\usepackage{caption}
\usepackage{subcaption}
\usepackage{hyperref} 
\usepackage{amsmath}
\usepackage{tikz}
\usetikzlibrary{tikzmark}
\usepackage{flushend}

\usepackage{amssymb}

\journal{}

\begin{document}\sloppy

\begin{frontmatter}



\title{Portability and Scalability Evaluation of Large-Scale Statistical Modeling and Prediction Software through HPC-Ready Containers}


\author[inst1]{Sameh Abdulah}

\affiliation[inst1]{organization={Extreme Computing Research Center (ECRC), King Abdullah University of Science and Technology (KAUST)},
            city={Thuwal},
            country={Saudi Arabia.}}

\author[inst2]{Jorge Ejarque}
\affiliation[inst2]{organization={Barcelona Supercomputing Center (BSC)},
            city={Barcelona},
            country={Spain.}}
            
\affiliation[inst3]{organization={Brightskies Inc.},
            city={Alexandria Governorate},
            country={Egypt.}}
\author[inst3] {Omar Marzouk}
\author[inst1] {Hatem Ltaief}
\author[inst1] {Ying Sun}
\author[inst1]{Marc G. Genton}
\author[inst2] {Rosa M. Badia}
\author[inst1]{David E. Keyes}

\begin{abstract}
HPC-based applications often have complex workflows with many software dependencies that hinder their portability on contemporary HPC architectures. In addition, these applications often require extraordinary efforts to deploy and execute at performance potential on new HPC systems, while the users expert in these applications generally have less expertise in HPC and related technologies. This paper provides a dynamic solution that facilitates containerization for transferring HPC software onto diverse parallel systems. The study relies on the HPC Workflow as a Service (HPCWaaS) paradigm proposed by the EuroHPC eFlows4HPC project. It offers to deploy workflows through containers tailored for any of a number of specific HPC systems. Traditional container image creation tools rely on OS system packages compiled for generic architecture families (x86\_64, amd64, ppc64, ...) and specific MPI or GPU runtime library versions. The containerization solution proposed in this paper leverages HPC Builders such as Spack or Easybuild and multi-platform builders such as buildx to create a service for automating the creation of container images for the software specific to each hardware architecture, aiming to sustain the overall performance of the software. We assess the efficiency of our proposed solution for porting the geostatistics ExaGeoStat software on various parallel systems while preserving the computational performance. The results show that the performance of the generated images is comparable with the native execution of the software on the same architectures. On the distributed-memory system, the containerized version can scale up to 256 nodes without impacting performance.
\end{abstract}




\end{frontmatter}


\section{Introduction}
Scientific computing has experienced a drastic shift from relying on single-core machines to today's massively parallel distributed-shared memory systems. High-Performance Computing (HPC) tools have added tremendous value to many applications, allowing them to handle larger workloads by providing more computing power and memory capacity. However, HPC systems have become increasingly more complicated and heterogeneous due to the large variety of hardware architectures and accelerators they incorporate. Moreover, the software components of these systems, including operating systems, programming languages, compilers, libraries, and tools, exhibit wide variation. This variety creates challenges for existing HPC software, including poor portability and low reusability. An extra virtualization layer is a natural way to tackle these challenges and allow easy and efficient portability of HPC-oriented software. The primary objective of this layer is to isolate the software stack from the underlying hardware platform and help domain scientists run HPC-oriented scientific applications without fully understanding how to optimize these applications on their HPC hardware. When the isolation happens at the operating system level, and the application can run in an isolated user space regardless of the underlying hardware, this is called ``containerization.''

Container systems operate by utilizing a shared operating system kernel across multiple container images. Each has a separate and isolated user-space environment, allowing many instances to run on the same host without interfering with each other. This technology appeared first to manage user access privileges inside the Unix operating system and gained popularity with the release of Docker~\cite{boettiger2015introduction} and its interaction with the cloud computing community in 2014. Containers have become popular in the cloud computing community due to their impact on improving the portability of the code and the small overhead they add to the execution compared to other virtual machine environments such as KVM~\cite{habib2008virtualization}, VMware Fusion~\cite{li2010selecting}, Hyper-V~\cite{djordjevic2021performance},  and Xen~\cite{abels2005overview}. Recently, containers have gained more attention in the HPC community, where several studies provide performance assessments on the benefits of containers in improving the scalability and portability of the HPC software. The studies argue that containers can be a ``golden solution'' for avoiding the complexity of the current HPC software setup and can guarantee portability on contemporary and future HPC systems. However, the performance achieved by containers relies on how the container images have been created. To achieve the best performance on an HPC system, applications must be installed taking into account the features of the system, such as the computer architecture, the available networking fabric, and/or the available accelerators. 

Traditionally, container images are created using OS packages that are binary installations pre-compiled with generic optimization flags, and communication libraries are compiled for generic networking fabric available in most computers. These installations are usable in all the processors of a family. Still, they cannot use the features of the more advanced processors, such as vector instructions and other compile-time optimizations. So, container images created with this methodology are more portable but can not achieve the best performance available in the system.

Trying to close the gap between deployment simplicity and performance, the eFlows4HPC project\footnote{https://eflows4hpc.eu} proposes a methodology to leverage containers to simplify the installation of complex workflows in HPC infrastructures while maintaining the performance at the same level as native installations. This paper proposes a service to automate the creation of container images tailored to specific HPC systems. From a generic recipe and a target system description, the service automates the creation of an image for this system, leveraging HPC package builders such as Spack~\cite{gamblin2015spack} and multi-platform container builders such as buildx, which can customize the building process for the HPC system features. The suggested approach enables the one-time preparation of installation scripts for all software dependencies using Spack, followed by the creation of container images for all necessary architectures.

We assess the performance of the proposed solution by porting the geospatial statistics application  \textit{ExaGeoStat}~\cite{abdulah2018exageostat}. \textit{ExaGeoStat} is a parallel software library for modeling spatial and spatio-temporal data through the Maximum Likelihood Estimation (MLE) method. Moreover, the software can predict missing geospatial data in space or space-time with the help of observed data and a set of obtained statistical parameters from the modeling process. In~\cite{abdulah2022large}, the authors propose \textit{ExaGeoStatR}, an {\em R} package that wraps the \textit{ExaGeoStat} software under the {\em R} environment. Statisticians can use \textit{ExaGeoStat} and  \textit{ExaGeoStatR} to tackle many operations on large geospatial data volumes if the complexity of installation does not limit its adoption.


In this work, we present the experience of using containers with the \textit{ExaGeoStat} software by building  HPC-ready containers for many architectures from a generic builder recipe. The assessment relies on pre-prepared Spack scripts to generate HPC-ready Singularity containers. The performance obtained in the experimentation demonstrates that containers can preserve the accuracy of shared-memory systems from various vendors and distributed memory systems under different conditions. The assessment includes two computation variants from \textit{ExaGeoStat}, including dense and Tile Low-Rank (TLR) approximation. 
The experiments also implicitly assess the performance of two commonly used parallel linear algebra libraries, i.e., Chameleon~\cite{chameleon}, HiCMA~\cite{abdulah2019hierarchical}, and the popular runtime system StarPU~\cite{augonnet2009starpu}.

\section{Contributions}
Our contributions can be summarized as follows:
\begin{itemize}
    \item We present an automated approach for generating HPC-ready containers for HPC software, ensuring portability across various hardware architectures while maintaining performance.
    \item We rely on the combined usage of Spack, Singularity, and Docker buildx in our container creation process. We employ a straightforward bash script to streamline the process of retrieving the generated containers on a specific hardware system.
    \item We comprehensively explain the pipeline involved in creating HPC-ready containers that can be easily applied to existing HPC software with minimal modifications.
    \item We employ the \textit{ExaGeoStat} software and its dependencies as an example to evaluate and analyze the effectiveness of the created containers. While our focus is primarily on the \textit{ExaGeoStat} software, it is important to note that our HPC-ready containers can be applied to a wide range of existing HPC software, provided they are compatible with the targeted hardware.
    \item We use multiple shared-memory systems and a distributed-shared-memory system to demonstrate the performance of our generated containers compared to the bare-metal installation of \textit{ExaGeoStat}. This enables us to assess the efficiency and effectiveness of the containers in various computing environments.
\end{itemize}

The paper is organized as follows: Sections~\ref{sec:related} and \ref{sec:background} present related work and background technologies, Section~\ref{sec:eflows} describes the eFlows4HPC project, and Section~\ref{sec:exageostat} gives an overview of the \textit{ExaGeoStat} software and its components. Section~\ref{sec:hpc_ready_cont} presents the methodology for generating HPC-Ready containers. Section~\ref{sec:experiment} describes extensive experimentation performed to evaluate the multiple HPC-ready containers, and Section~\ref{sec:conclus} summarizes the work.

 \section{Related Work}  
\label{sec:related}
Many studies aim to evaluate the efficiency of VMs and containers with HPC applications and systems. For instance, in~\cite{youseff2006evaluating}, an evaluation study on the Xen virtualization system for the HPC systems has been shown. The Xen virtualization system is a technique that allows running separate operating systems on the same hardware~\cite{abels2005overview}. Xen is popular for its high performance, scalability, and reliability. The authors prove that Xen virtualization can be a practical solution for the HPC software portability with 2-3\% overhead compared to executing the original kernels. Another study example comparing three VM solutions has been proposed in~\cite{walters2008comparison}. Among VMWare Server, Xen, and OpenVZ, OpenVZ provides the most satisfactory performance for HPC applications—other studies in the virtualization solution are~\cite{matthews2007quantifying,mergen2006virtualization,somani2009application,xavier2013performance}.

Containers were proposed through the release of Docker~\cite{boettiger2015introduction} in 2013. Containerization has gained more attention in software portability than VMs since it depends on virtualizing the software layer, not the entire machine, where different instances share the same host OS kernel. Due to its ability to isolate the OS and hardware layer from the software layer, it is being adopted in the HPC community to port complex applications. Many studies have evaluated the efficiency of containers in porting HPC software. For instance, the authors of~\cite{rudyy2019containers} provide a scalability and portability study for a production biological simulation application that relies on containers. The study embraces three container systems, Docker, Singularity~\cite{kurtzer2017singularity}, and Shifter~\cite{gerhardt2017shifter}, and claims that Singularity can provide the scalability and portability performance required by the target application. Another study ~\cite{ruiz2015performance} includes performance evaluation of containers under heavy HPC workloads. The study also assesses the efficiency of container-based HPC software with different Linux kernels on HPC systems. Other studies evaluating containers in HPC contexts are~\cite{abraham2020use,gantikow2020rootless,higgins2015orchestrating,torrez2019hpc}.

Although containers have been used to port HPC applications for several years, all the efforts were made individually and for a specific application or software. The HPC community can benefit from a standard workflow manager that can help simplify the porting process for different HPC applications. In~\cite{EJARQUE2022414}, the HPC Workflow as a Services (HPCWaaS) paradigm has been proposed under the eFlows4HPC project to enable encapsulating the HPC software workflow more dynamically and intelligently. The new paradigm adopted the concept of ``HPC-ready containers'' to build efficient and more reliable containers for contemporary HPC applications. In this paper, we assess the HPC-ready containers provided by the HPCWaaS paradigm in the eFlows4HPC project using the \textit{ExaGeoStat} software. The selected software mainly relies on dense linear algebra operations through exact and low-rank approximation computations to a covariance matrix that can reach millions by millions in size~\cite{cao2022reshaping}. \textit{ExaGeoStat} can be executed in all leading-edge parallel hardware architectures, including accelerators which motivates us to use it as a test case in our study. More details about \textit{ExaGeoStat} can be found on~\cite{abdulah2018exageostat,abdulah2018parallel}.

\section{Background}
\label{sec:background}
The following subsections provide an overview of the Containerization concept, Spack package manager, and Singularity. Spack and Singularity are the two main components of the HPC-ready containers in the eFlows4HPC project.

\subsection{Containerization}
Performance portability of HPC software on different HPC systems is difficult because of the complicated structure of its dependencies. The search for a robust solution allowing fast and secure porting on HPC systems led to containerization, which aims to package the software code and the required operating system libraries and dependencies into a single executable file that can run on different HPC systems while preserving the expected performance on these systems. Furthermore, containers are more robust than Virtual Machines (VMs) solutions since containers isolate the software layer from the hardware layer through the underlying operating system namespaces, which places the software much closer to the physical system, thus improving performance~\cite{rudyy2019containers}. Docker~\cite{boettiger2015introduction}, Singularity~\cite{kurtzer2017singularity}, and Shifter~\cite{gerhardt2017shifter} are three prime examples.

 \subsection{Spack Package Manager}
The Spack package manager~\cite{gamblin2015spack} provides a novel, recursive specification syntax to invoke parametric builds of packages
and dependencies. It allows any number of builds to coexist on
the same system, and it ensures that installed packages can find
their dependencies and libraries, regardless of the environment. Spack provides a simple spec syntax, allowing users to configure versions and options easily and precisely. It simplifies the packaging job for authors as the package files are written in Python. Also, the specs allow them to maintain a single file for different package builds.

Spack provides Spack Environments for the creation of container images. An environment is used to group together a set of specs for the purpose of building, rebuilding, and deploying coherently. Environments separate the steps of (a) choosing what to install, (b) concretizing, and (c) installing. This allows Environments to remain stable and repeatable. Also, environments allow several specs to be built at once. In addition, an Environment that is built as a whole can be loaded as a whole into the user environment. The environment to be created is defined in a manifest file in YAML format (spack.yaml). 

 \subsection{Singularity Container System} 
The appearance of container solutions such as Docker\footnote{Docker website, https://www.docker.com} offered improvements over standard virtual machines. However, for the scientific world, and specifically HPC communities, the Docker technology does not fit cleanly. The installation of Docker on  HPC environments would mean a level of security risk deemed unreasonable, preventing it from being embraced by a large community. This security risk comes from Docker's container processes spawned as children of a root-owned Docker daemon. 

Singularity~\cite{kurtzer2017singularity} is a container solution created by
the necessity for scientific application-driven workloads. 
Singularity containers can be built on the user's laptop and run on many large HPC clusters and in single servers or the cloud. 
Singularity offers mobility by enabling environments to be ported via a single image file and is designed with the features necessary to allow seamless integration with any scientific computational resources. Singularity is a pioneer in providing an accessible environment for users and administrators since it was developed in collaboration with HPC administrators, developers, and research scientists.

The goal of Singularity is to support existing and traditional HPC resources as easily as
installing a single package onto the host operating system. 
Singularity can run on a large number of Linux distributions, including vintage ones. 
Singularity natively supports technologies such as InfiniBand
and Lustre while at the same time integrating seamlessly with any resource manager (e.g.,
SLURM, Torque, SGE, etc.) as a result of the fact that Singularity is run like any other command
on the system. Singularity also includes a SLURM plugin that allows SLURM
jobs to be natively run within a Singularity container. Singularity also enables access to unique
host resources such as GPUs. 
Regarding security issues, Singularity does not provide the ability to escalate
permission inside a container. With Singularity, if a user does not have root access on the target system, the user cannot escalate privileges within the container to root either.
 
\section{The eFlows4HPC Project}
\label{sec:eflows}

The EuroHPC Joint Undertaking (EuroHPC JU) aims to develop a world-class supercomputing ecosystem in Europe.  It is procuring and deploying five petascale and three pre-exascale systems in Europe and has announced the installation of the first European Exascale system in 2023. These systems will be capable of running large and complex applications, which in many cases will not be exclusively traditional simulation workloads but include aspects of artificial intelligence and data analytics. The EuroHPC JU, in addition, is funding projects co-designing new systems and the necessary software stack and applications that will leverage the future exascale systems. 

The eFlows4HPC~\cite{EJARQUE2022414} is a EuroHPC project focusing on developing workflows. eFlows4HPC aims to provide a software stack that simplifies the development of workflows that combine HPC simulation and modeling with artificial intelligence and data analytics. The project also seeks to enable the development of workflows that can react to external or internal events (i.e., changes in input data or code exceptions) and dynamically change the behavior of the workflow. In addition, the eFlows4HPC software stack provides runtimes that execute the workflows efficiently in terms of time and energy. The project is also developing the concept of the HPC Workflows as a Service (HPCWaaS) concept, which aims at providing a mechanism to make the use and reuse of the workflows in HPC infrastructures easier. With this goal, the project has designed the HPCWaaS interface, which supports all the phases of the workflow lifetime: workflow development, deployment, credential management, and execution. 

This paper leverages the deployment methodologies developed in the eFlows4HPC project, which are based on a service that can generate HPC-ready containers. The container greatly simplifies the software deployment phase in any computing infrastructure. However, if a generic container image is used in an HPC system, this image will probably not benefit from the specific hardware features and specialized libraries of the system. Taking this into account, the project has designed a methodology that enables the automatic generation of containers that leverage the hardware and software features of the specific system. 

The methodology used in this article is based on the Spack package manager. The application developer provides a YAML file with the software requirements and a Python script that includes the package development procedure in Spack to build the specific HPC-ready container. The container generation service uses these files and the information about the target system for which we would like to generate the container image. Once the image for a specific system has been created, it can be stored in an image repository for reuse.

\section{ExaGeoStat: The Use Case Software}
\label{sec:exageostat}
Spatial statistics methods aim to model spatial and spatio-temporal data to obtain a set of statistical parameters that can be later used to predict missing values in space or space-time. Geospatial data can be found in many disciplines, including climatology, topography, and geology. Indeed, traditional geostatistical tools are readily overwhelmed with the huge amount of geospatial data coming from sources such as satellites, and novel solutions are needed to handle these large data volumes. \textit{ExaGeoStat}~\cite{abdulah2018exageostat} is high-performance software for computational geostatistics on many-core systems for tackling large data volumes through state-of-the-art HPC-based solutions that has run on numerous Top100 systems of different architectures and was selected as a 2022 Gordon Bell finalist~\cite{cao2022reshaping}. 

\subsection{ExaGeoStat Software Operations}
\begin{figure*}[h!]
\begin{center}
\includegraphics[width=0.8\linewidth]{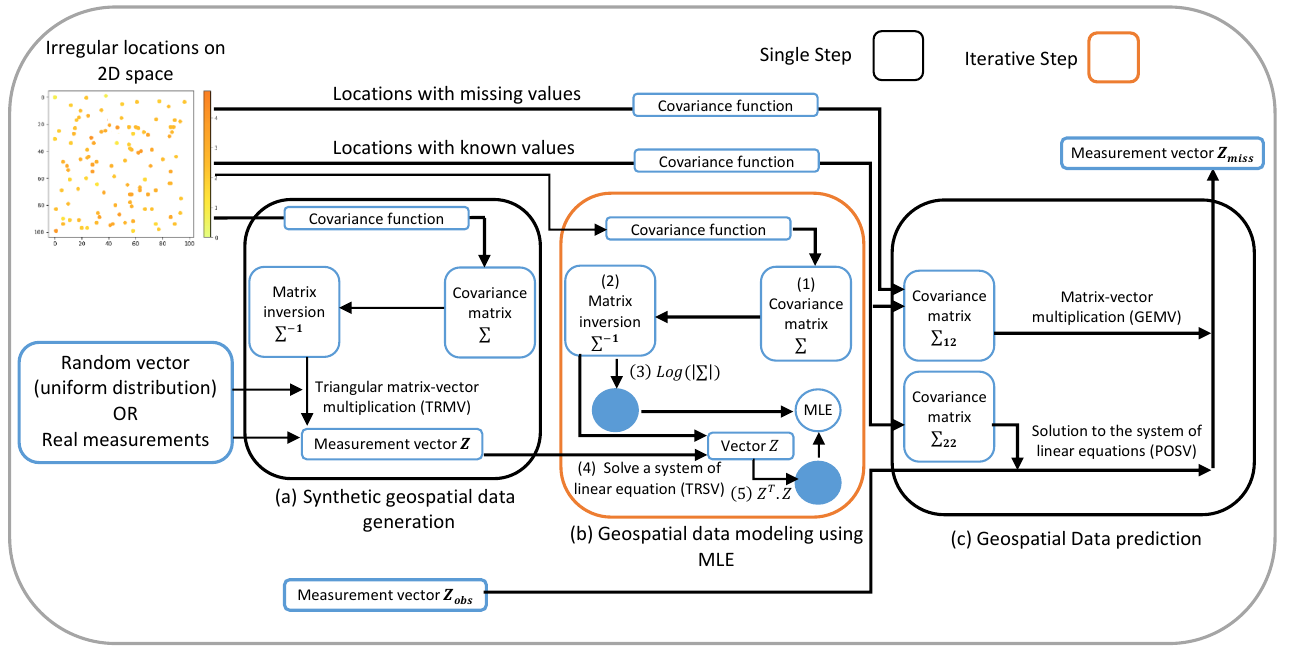}
\caption{\textit{ExaGeoStat} main operations: (a) Synthetic geospatial data generation (b) Geospatial data modeling using Maximum Likelihood Estimation (MLE) (c) Geospatial data prediction.}
\label{fig:libs}
\end{center}
\end{figure*}

\textit{ExaGeoStat} offers three primary geostatistical operations: synthetic geospatial data generation, geospatial data modeling using Maximum Likelihood Estimation (MLE), and geospatial data prediction through kriging. Figure~\ref{fig:libs} explains the operations involved in each component.

\subsubsection{Synthetic Geospatial Data Generation:}
\textit{ExaGeoStat} provides an internal data generator of geospatial data for experiments under prescribed conditions such as distribution shape and correlation lengths in space and time. For illustrations herein, which focus on performance portability and not statistical features, we consider only the 2D spatial Gaussian distribution. Assuming $n$ spatial locations uniformly distributed in the unit square, the covariance matrix $\mathbf{\Sigma(\boldsymbol \theta)}$ can be built using a wide range of covariance functions, including the Matérn covariance function~\cite{abdulah2018exageostat}. The first task is to compute the inverse of $\mathbf{\Sigma(\boldsymbol \theta)}$  using Cholesky factorization: $\mathbf{\Sigma(\boldsymbol \theta)} = \mathbf{L}\cdot \mathbf{L}^\top$. The second task is to compute the measurement vector $\mathbf{Z}$ as $\mathbf{Z =L\cdot e}$, where $\mathbf{e}$ is another random standard Gaussian distributed vector. The geospatial data are represented by the set of locations and corresponding measurements in each location, $\mathbf{Z }$. The entire operation is shown by Figure~\ref{fig:libs} (a).

\subsubsection{Geospatial Data Modeling:}
The Maximum Likelihood Estimation (MLE) method is used for geospatial data modeling in \textit{ExaGeoStat}. The geospatial data are usually modeled in spatial statistics as a realization from a Gaussian spatial random field. Assuming $\mathbf{Z}$ vector be a realization of a Gaussian random field $\mathbf{Z(s)}$ where $\bf s$ is the set of spatial locations, $\mathbf{\Sigma(\boldsymbol \theta)}$ is a covariance matrix  with dimension $n$, and the random field $\mathbf{Z}$ has a mean zero, the statistical parameter vector $\boldsymbol \theta$ can be obtained based on the Gaussian log-likelihood function:
\begin{equation}
	\label{eq:likeli}
	\ell(\boldsymbol \theta)=-\frac{n}{2}\log(2\pi) - \frac{1}{2}\log |{{\mathbf{ \Sigma}}({\boldsymbol \theta})}|-\frac{1}{2}{\bf Z}^\top {\mathbf{\Sigma}}({\boldsymbol \theta})^{-1}{\mathbf{Z}}.
\end{equation}

Figure~\ref{fig:libs} (b) explains in detail how to compute the likelihood function in five steps: 1) generate the covariance function $\mathbf{\Sigma(\boldsymbol \theta)}$ from an initial parameter vector $\boldsymbol \theta_{init}$; 2) Compute the inverse of $\mathbf{\Sigma(\boldsymbol \theta)}$ using the Cholesky factorization operation $\mathbf{\Sigma(\boldsymbol \theta)}=\mathbf{L \cdot L}^\top$; 3) Compute the log determinant of $\mathbf{\Sigma(\boldsymbol \theta)}$; 4) Solve the system of linear equations $\mathbf{L}^{-1}{\bf Z}$;  5) Compute the dot product ${\bf Z}^\top\cdot {\bf Z}$. The constant expression $c=-\frac{n}{2}\log(2\pi)$ can be ignored since it does not affect the optimization problem.

\subsubsection{Geospatial Data Prediction:}
Data modeling using MLE produces a set of statistical parameters $(\hat{\boldsymbol \theta}$ that can be used in predicting missing values. Assuming unknown values $\mathbf{Z}_1$ with size $m$ and known values $\mathbf{Z}_2$ with size $n$, and $\mathbf{Z}_2$ has a zero-mean function, so $\mathbf{Z}_1$ can be obtained as

\begin{equation}
	\label{eq:pred}
\mathbf{Z}_1=\mathbf{\Sigma}_{12}\mathbf{\Sigma}_{22}^{-1}\mathbf{Z}_2,
\end{equation}
where $\mathbf{\Sigma}_{12}$ is the covariance matrix generated from the distance matrix between the locations of the missing values and the observed values, $\mathbf{\Sigma}_{22}$ is the covariance matrix generated from the distance matrix between the locations of the observed values. The proof and the explanation of equation~(\ref{eq:pred}) can be found on~\cite{genton2007separable}. Figure~\ref{fig:libs}(c) explains the prediction operation in detail.

\subsection{Parallel Linear Algebra Mathematics in ExaGeoStat}
\textit{ExaGeoStat} enables parallel computation of the MLE operation through the state-of-the-art parallel linear algebra libraries, Chameleon for dense computation, and HiCMA for Tile Low-Rank (TLR) approximation~\cite{abdulah2019hierarchical}. Both libraries rely on the StarPU runtime system to enable porting the code into different parallel hardware systems, including GPUs. Through tile-based algorithms, Chameleon and HiCMA performed better on shared-memory and distributed-memory systems compared to block-based algorithms in LAPACK~\cite{anderson1999lapack} and ScaLAPACK~\cite{choi1992scalapack}. In tile-based algorithms, matrix tiles are distributed to different processing units to allow faster execution of the underlying linear algebra operations that can accelerate applications that mainly rely on linear algebra such as \textit{ExaGeoStat}~\cite{abdulah2018exageostat}.

In HiCMA, low-rank approximation has been used in each matrix tile with a certain accuracy by compressing each tile using Singular Value Decomposition (SVD). SVD can be replaced by
Randomized SVD (RSVD), or Adaptive Cross Approximation (ACA), to approximate each off-diagonal tile up to a certain user-defined accuracy threshold. Using SVD, each tile $(i, j)$ in the covariance matrix can be represented by the product of two matrices $\mathbf{U}_{ij}$ and $\mathbf{V}_{ij}$,
with a size of $nb$ × $k$, where $nb$ represents the tile size which is a tunable parameter that has an impact on the overall performance. TLR approximation can compress the whole covariance matrix and improve the performance of executing linear algebra operations on large-scale systems. More information about TLR \textit{ExaGeoStat} can be found in~\cite{abdulah2018parallel}.

\section{HPC-Ready Containers Creation Mechanism}
\label{sec:hpc_ready_cont}
Like many existing HPC software, \textit{ExaGeoStat} relies on several software dependencies to improve the code portability across various hardware architectures. \textit{ExaGeoStat} depends on several libraries, including BLAS, CUDA, MPI, NLOPT, HWLOC, GSL, StarPU, Chameleon, STARS-H, and HiCMA~\cite{abdulah2019hierarchical}. Therefore, using containers is essential to facilitating seamless code portability.

\begin{figure*}[h]
\begin{center}
\includegraphics[width=0.7\linewidth]{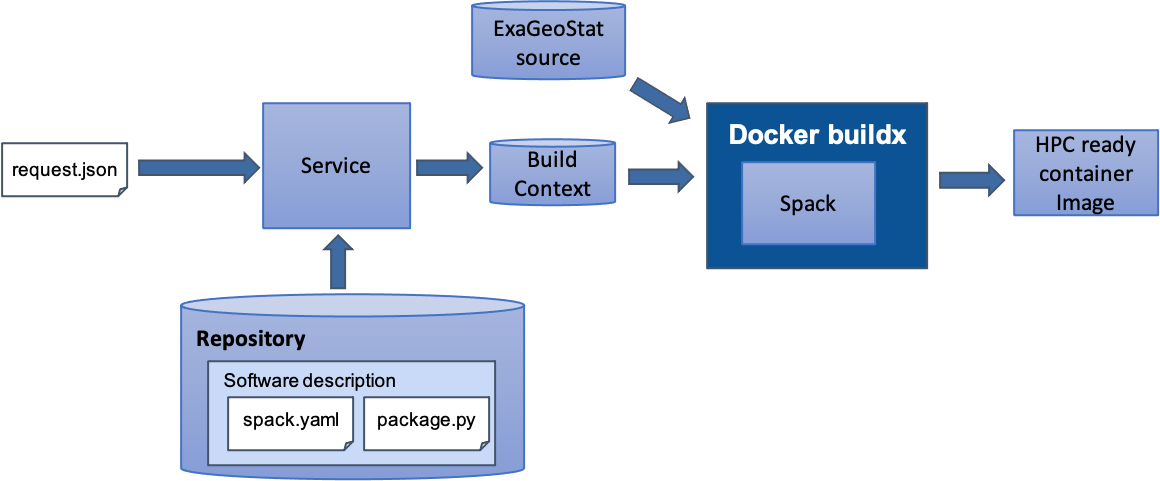}
\caption{Pipeline for the creation of HPC-ready containers.}
\label{fig:containers}
\end{center}
\end{figure*}

In this work, we propose generating containers for HPC software that can be easily created and adapted for different hardware architectures. In this section, we describe the proposed HPC-Ready containers in the context of \textit{ExaGeoStat} software. The eFlows4HPC project has proposed the Container Image Creation 
service\footnote{\url{https://github.com/eflows4hpc/image_creation}}, which aims at simplifying the creation of container images for HPC systems. This service leverages Docker buildx and Spack environments to automatize the generation of the containers for specific HPC platforms. It allows developers to use containers to facilitate the deployment in the HPC platforms. To run the software on a new hardware system, the developer simply needs to copy or pull the container image. This allows for easy execution without additional steps and achieves the same performance as a native installation. Figure~\ref{fig:containers} shows the complete pipeline the Container Image Creation service uses to create the HPC-ready containers. 

The repository contains a {\tt spack.yaml} Spack environment manifest file per application includes the software required to build an image for an HPC application. These manifest files outline the steps and dependencies needed to install the HPC application successfully. The repository also stores a generic {\tt package.py} that includes the installation description according to the Spack schemas. The developer only needs to write these two files once, as they can be reused to create containers for different HPC platforms. These files are generic, platform-agnostic, and must be customized to build the HPC-ready containers. As an example, for the \textit{ExaGeoStat} case, we store the Spack environment manifest file at 
\url{https://github.com/eflows4hpc/workflow-registry/tree/v2/kaust/exageostat/spack.yaml}
and Spack installation descriptions (package.py) for ExaGeoStat and its dependencies at
\url{https://github.com/eflows4hpc/software-catalog/tree/new_developments/packages/exageostat}.



\begin{figure}
\centering
\begin{minipage}{0.9\linewidth}
\includegraphics[width=1\linewidth]{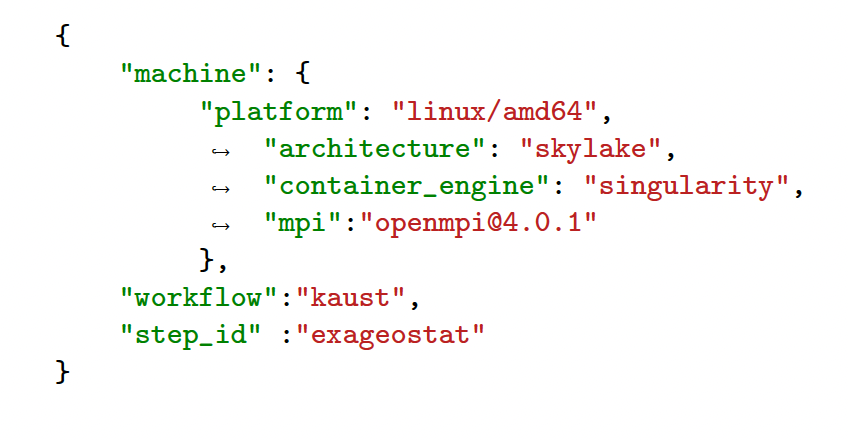}
  \caption{Container configuration (JSON file)}
  \label{fig:json}
\end{minipage}
\end{figure}

To manage the {\tt spack.yaml}  and the {\tt package.py} files, the developer should create a JSON document to build a container image for a specific HPC system. The JSON document describes the requested characteristics of the underlying hardware. As shown by the given example in figure~\ref{fig:json}, the JSON document has several fields. 
The field {\tt machine} specifies the target platform and architecture of the underlying system. In the example, two optional fields are specified to indicate that we are creating an image to be managed by Singularity and a specific {\tt MPI} version needed in that system. The fields {\tt workflow} and {\tt step\_id} are used to obtain the {\tt spack.yml} and the {\tt package.py} files in the repository. 

The service uses all this data to create a container build context. The service extends the generic Spack environment with the machine description to achieve this. So the extended environment will specify that required software packages will include the specific MPI and GPU runtimes and the CPU architecture available in the HPC system. With these extra definitions, we will ensure that the generated image will be compiled considering the processor features, and we will be able to access the MPI fabric and GPU accelerators. Apart from the extended environment, the build context will include the required Spack descriptions downloaded from the repository and a Dockerfile, which executes the Spack command to install the environment in the container image. 

After generating the build context, the service invokes the Docker buildx tool, passing the generated build context and the platform specified in the machine description. If the platform is not the same as the host running the service, the buildx tool will create an emulated environment to build the image. Within build, Spack is called with the corresponding options to build the specific image. In the case of \textit{ExaGeoStat}, the Spack build will download the source code of \textit{ExaGeoStat}, its dependencies, and the corresponding MPI and GPU runtime and install all of them according to the target HPC platform features.

After the build process, the container image is stored in the repository with accompanying metadata indicating the characteristics of the target machine. This is useful to detect if an image can be reused for another request, which requires the build the same application for an HPC platform with the same characteristics. Once it is stored, the image is converted to the format which is supported by the target machine. In the case of the example, the generated image is converted to a Singularity Image Format (SIF).

\begin{figure}[!tb]
\centering
\begin{minipage}{0.9\linewidth}
\includegraphics[width=1\linewidth]{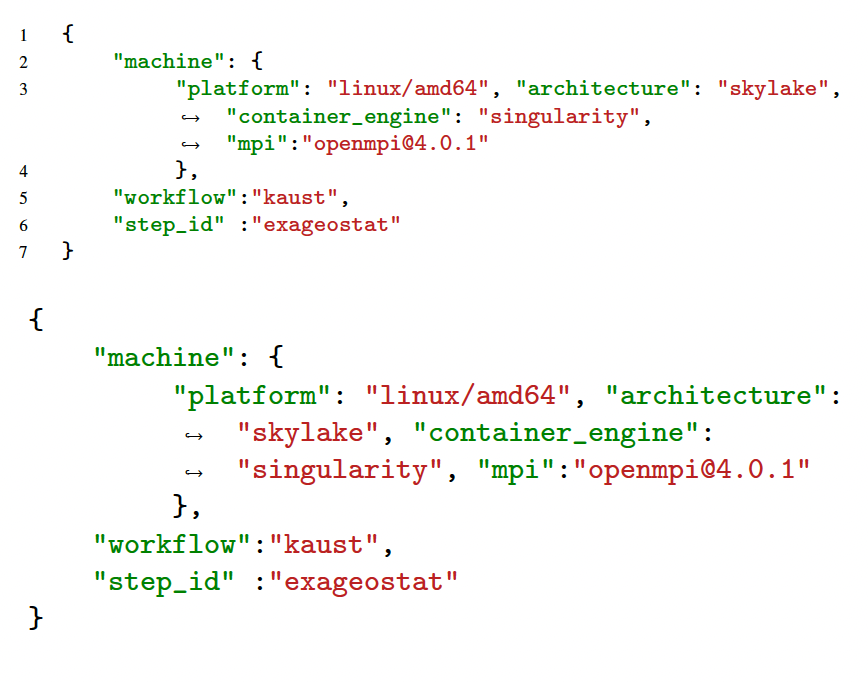}
  \caption{Client usage}
  \label{fig:usage}
\end{minipage}
\end{figure}

To manage the creation of the container images, the service provides an API with three methods to build, check the status, and download the image. To simplify the access to the service, we have developed a simple {\tt bash} client to invoke the service. Figure~\ref{fig:usage} shows the usage of this client.

\section {Experimental Evaluation } 

In this section, we assess the efficiency of the HPC-ready containers using the \textit{ExaGeoStat} software by conducting experiments in different parallel architectures and comparing the performance with the native build. We rely on shared-memory systems with/without accelerators and a distributed-memory system to evaluate the performance of the Singularity images under various workloads.

\label{sec:experiment}
\subsection{Experimental Testbed}
To assess the performance and the portability of the 
HPC-ready containers, we use a wide range of manycore
systems: a dual-socket 18-core Intel Haswell Intel Xeon E5-2699 running at 2.30GHz, 
a dual-socket 28-core Intel Skylake Intel Xeon Platinum 8176 running at 2.10 GHz, 
a dual-socket  32-core AMD Naples AMD EPYC 7601 running at 2.2GHz, 
a dual-socket 64-core AMD Milan AMD EPYC 7713, 
a dual-socket 28-core Intel Icelake Intel Xeon Gold 6330 CPU
running at 2.00GHz equipped with 4 NVIDIA A100 GPUs, 
a dual-socket 20-core Intel Skylake Intel Xeon Gold 6148 running at 2.40GHz equipped with 2 NVIDIA V100 GPUs. 
We also rely on MareNostrum 4 system, a Barcelona Supercomputing Center (BSC)  supercomputer with $3{,}456$ Lenovo ThinkSystem compute nodes where each node has two 24-core Intel Xeon Platinum chips (a total of $165,888$) processors and $390$ terabytes memory. In our experiments, we use up to 256 nodes on this system.

We rely on GCC v10.2.0 to compile and link against HWLOC v2.8.0, GSL v2.7.1, StarPU v1.3.9, Intel MKL v11.3.1, NLOPT v2.7.0, HiCMA v0.1.1, Chameleon v0.9.2, STARS-H v0.1.0, and ExaGeoStat v1.1.0.  For all the CPU-based architectures, the Singularity image size is about $2.2$ GB, the CPU/GPU-based image size is about $8.4$ GB, and the MPI-based (only CPU) image size is about $2.3$ GB.

All the performance results are an average of 10 different runs for the three primary operations in \textit{ExaGeoStat}: data generation, modeling, and prediction. In addition, the iterative modeling operation has been performed for 10 iterations. This mechanism allows better performance assessment to avoid any errors in the runs.

\subsection{Containers Portability on Shared-Memory Systems}
\begin{figure*}

          \begin{subfigure}[b]{\textwidth}
         \centering
         \includegraphics[width=0.26\textwidth]{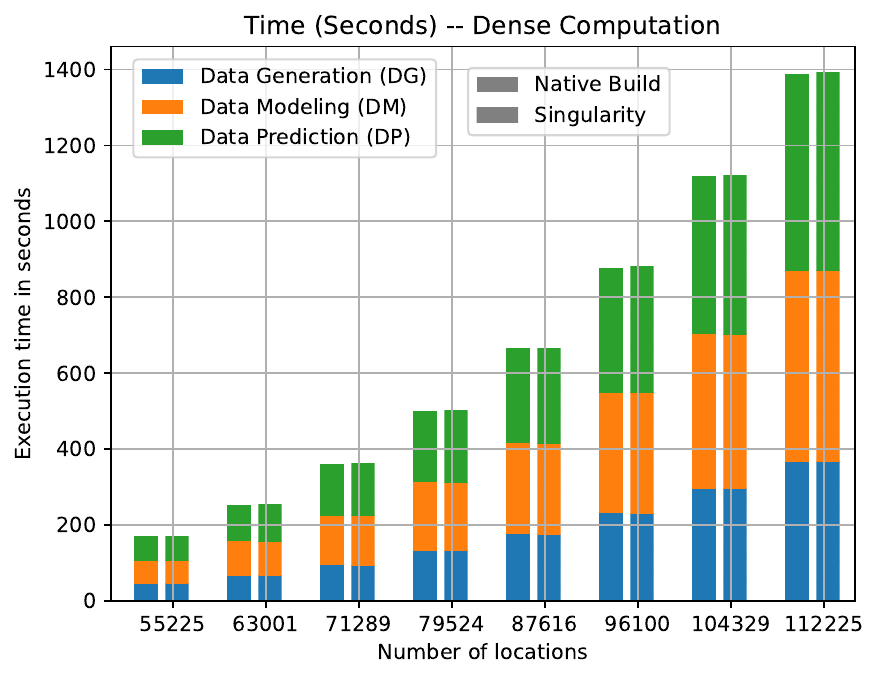}
                 \hspace{2mm}
         \includegraphics[width=0.26\textwidth]{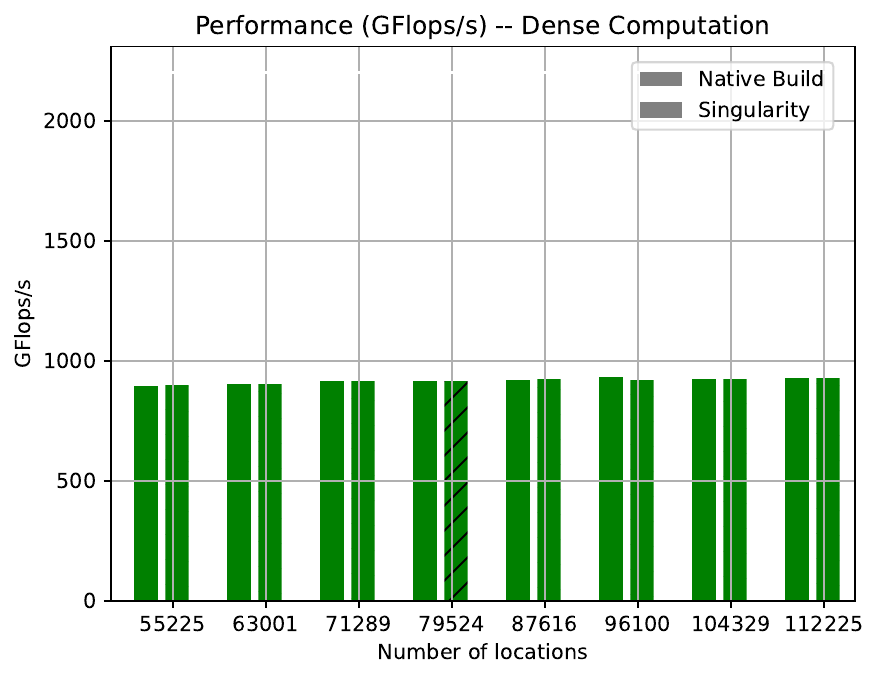}
                 \hspace{2mm}
         \includegraphics[width=0.26\textwidth]{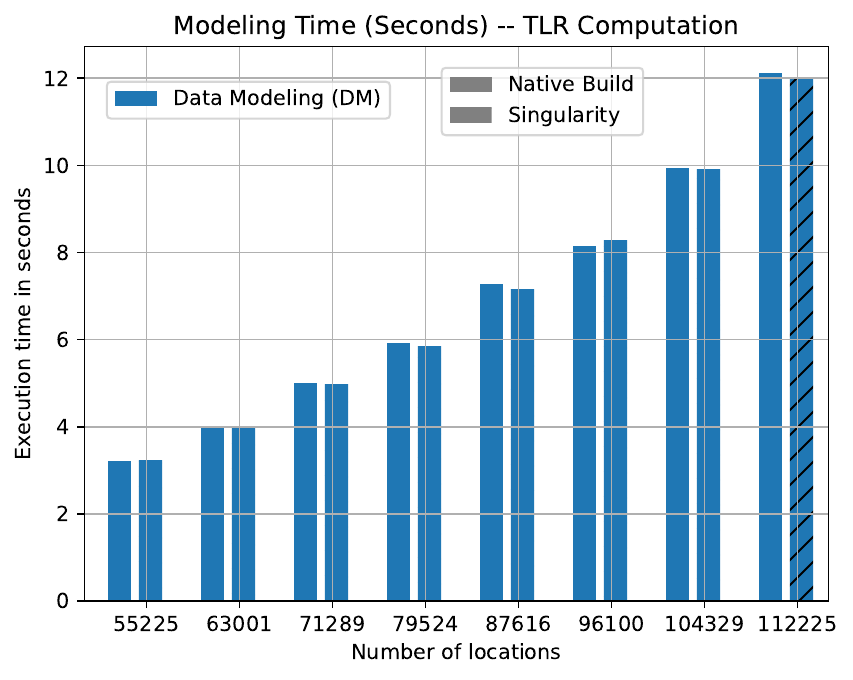}
         \caption{18-core Intel Haswell.}
         \label{fig:haswell}
     \end{subfigure}

        \begin{subfigure}[b]{\textwidth}
         \centering
         \includegraphics[width=0.26\textwidth]{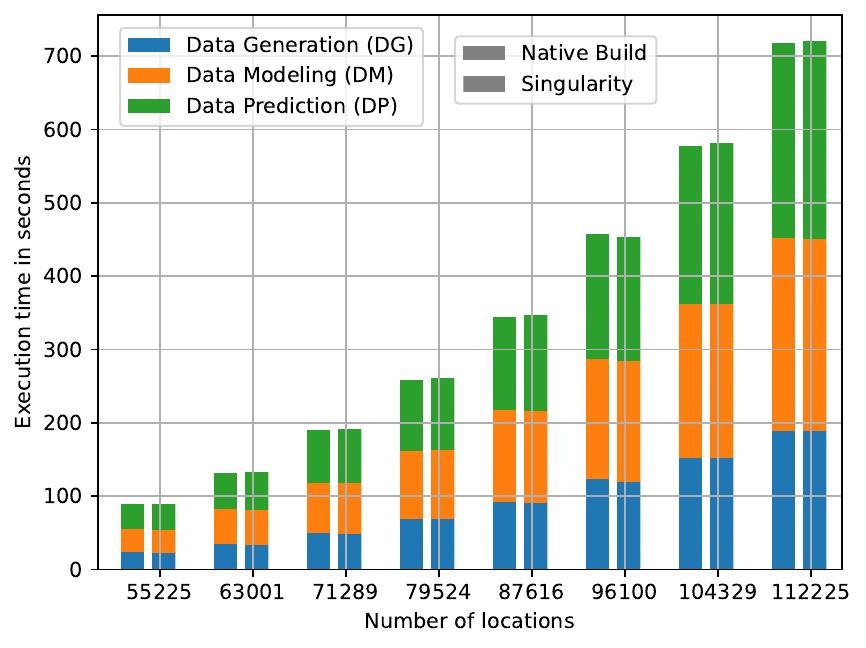}
                 \hspace{2mm}
         \includegraphics[width=0.26\textwidth]{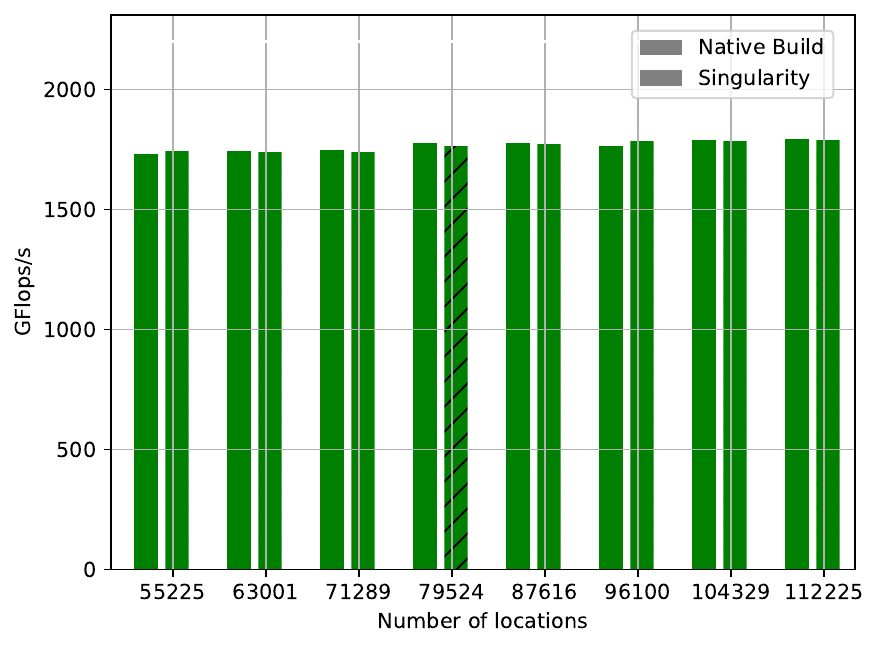}
                 \hspace{2mm}
         \includegraphics[width=0.26\textwidth]{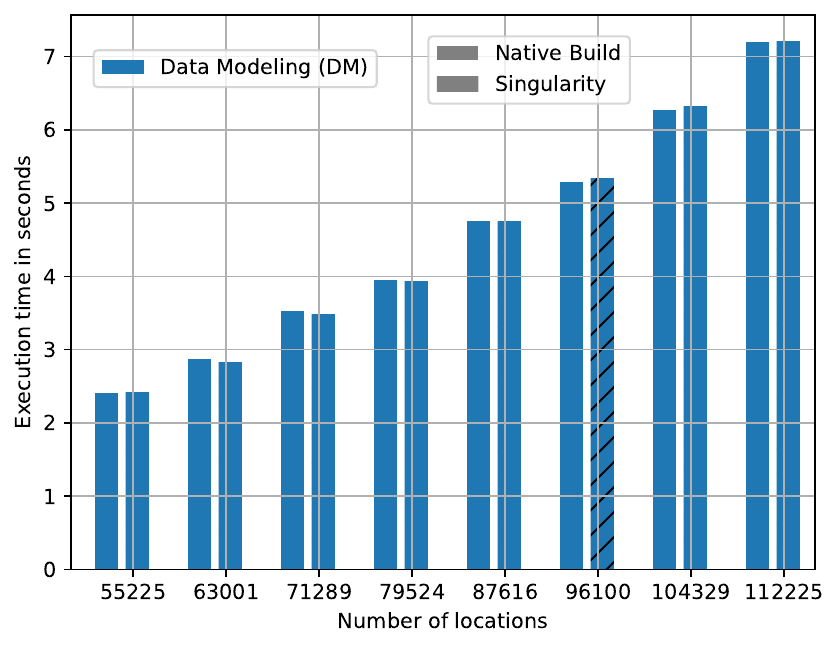}
         \caption{28-core Intel Skylake.} 
         \label{fig:skylake}
     \end{subfigure}

        \begin{subfigure}[b]{\textwidth}
         \centering
         \includegraphics[width=0.26\textwidth]{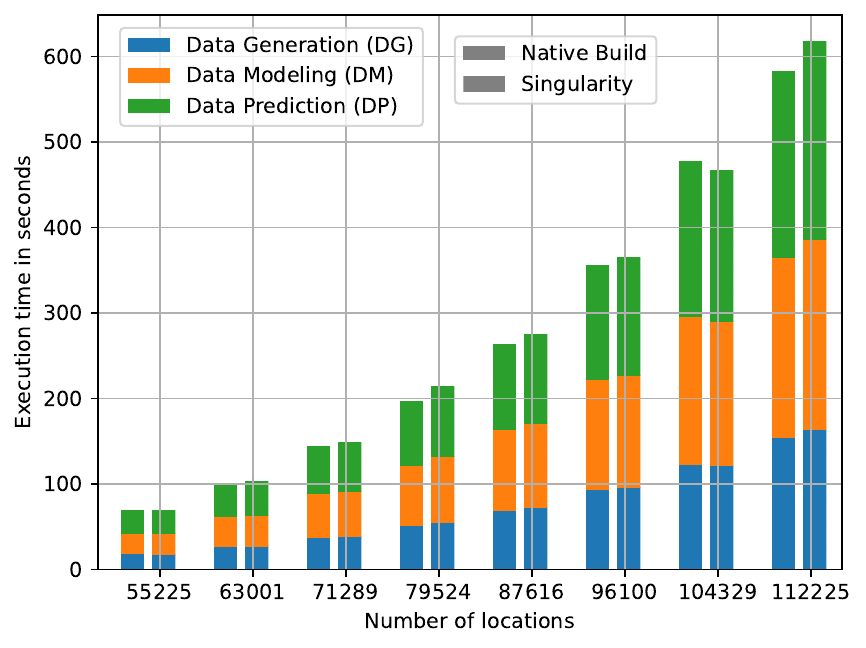}
                 \hspace{2mm}
         \includegraphics[width=0.26\textwidth]{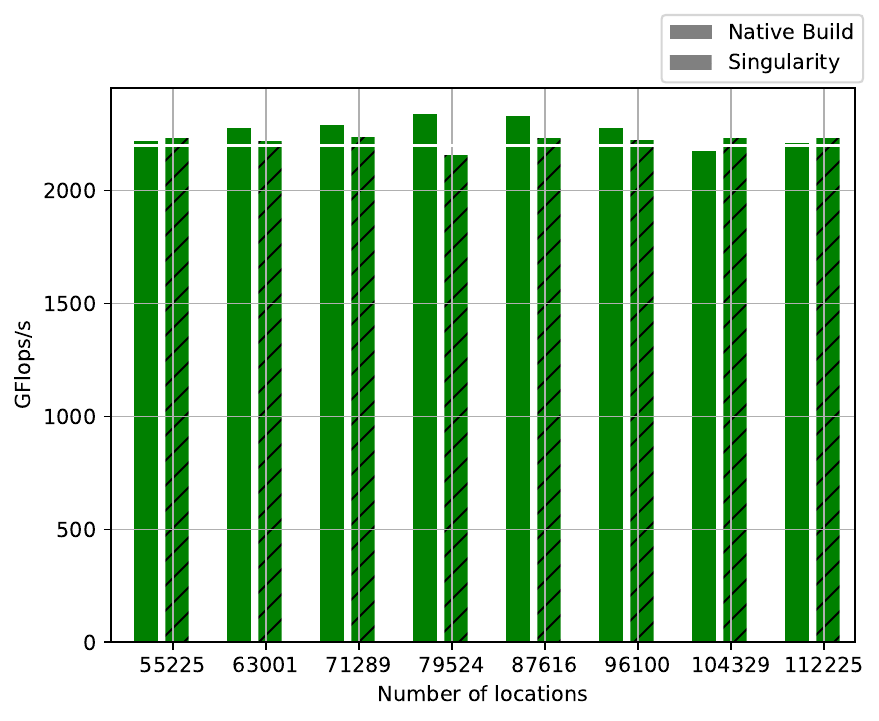}
                 \hspace{2mm}
         \includegraphics[width=0.26\textwidth]{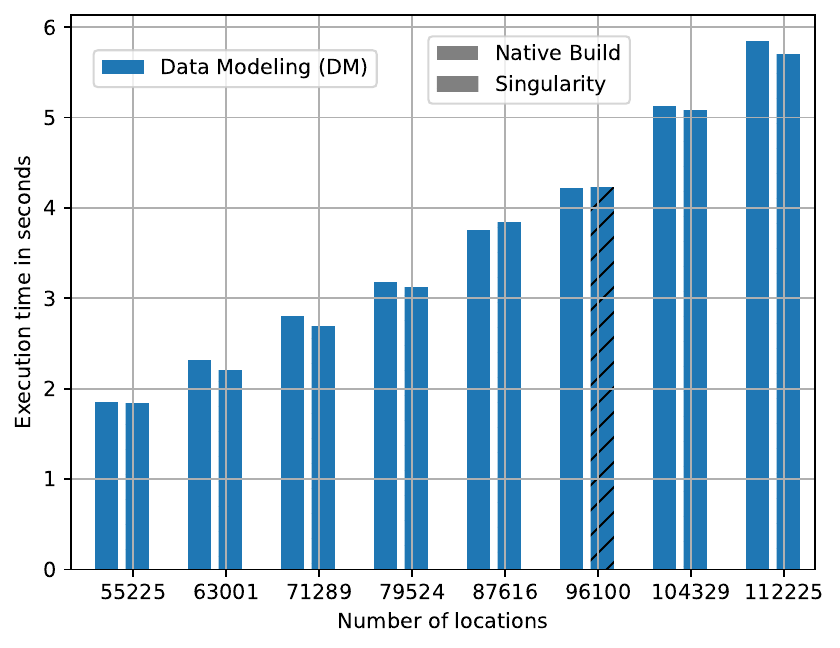}
         \caption{28-core Intel Icelake.}
         \label{fig:icelake}
     \end{subfigure}
             \begin{subfigure}[b]{\textwidth}
         \centering
         \includegraphics[width=0.26\textwidth]{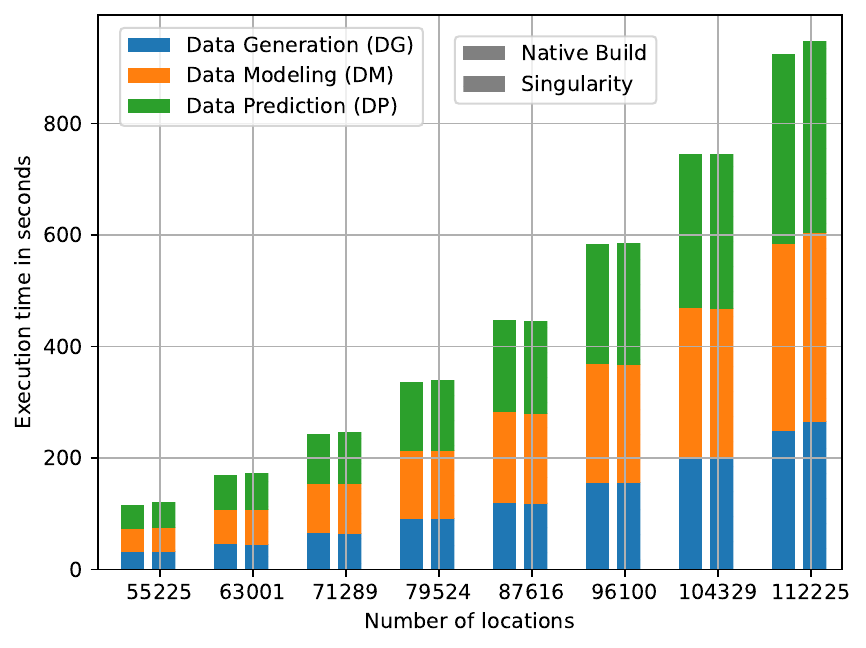}
                 \hspace{2mm}
         \includegraphics[width=0.26\textwidth]{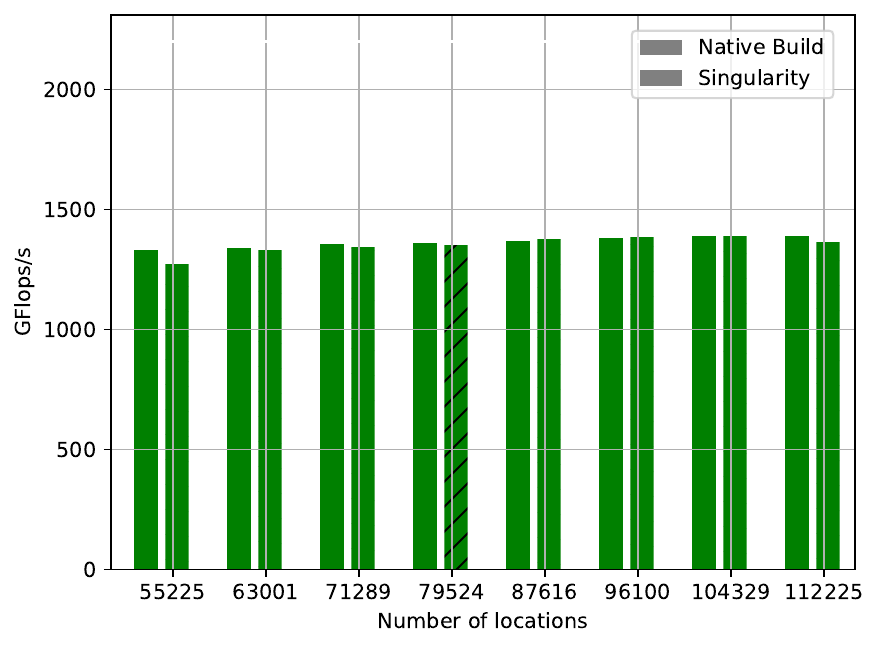}
                 \hspace{2mm}
         \includegraphics[width=0.26\textwidth]{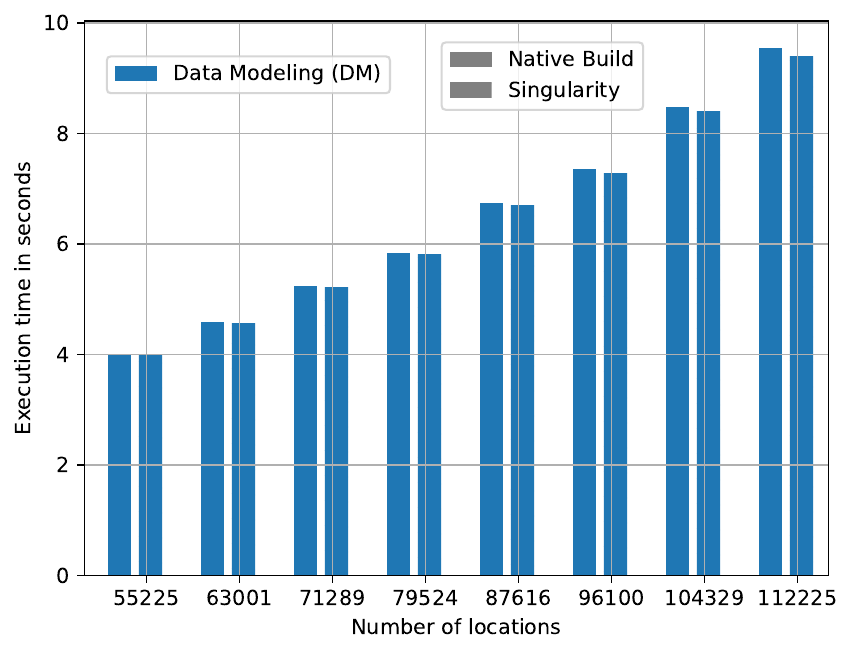}
         \caption{64-core AMD Milan.}
         \label{fig:milan}
     \end{subfigure}
        \begin{subfigure}[b]{\textwidth}
         \centering
         \includegraphics[width=0.26\textwidth]{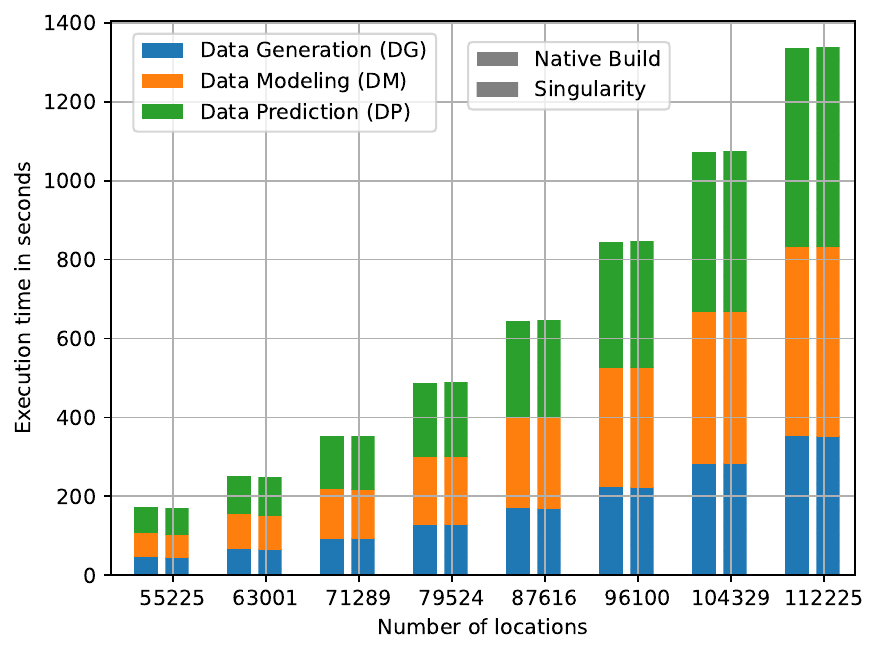}
         \hspace{2mm}
         \includegraphics[width=0.26\textwidth]{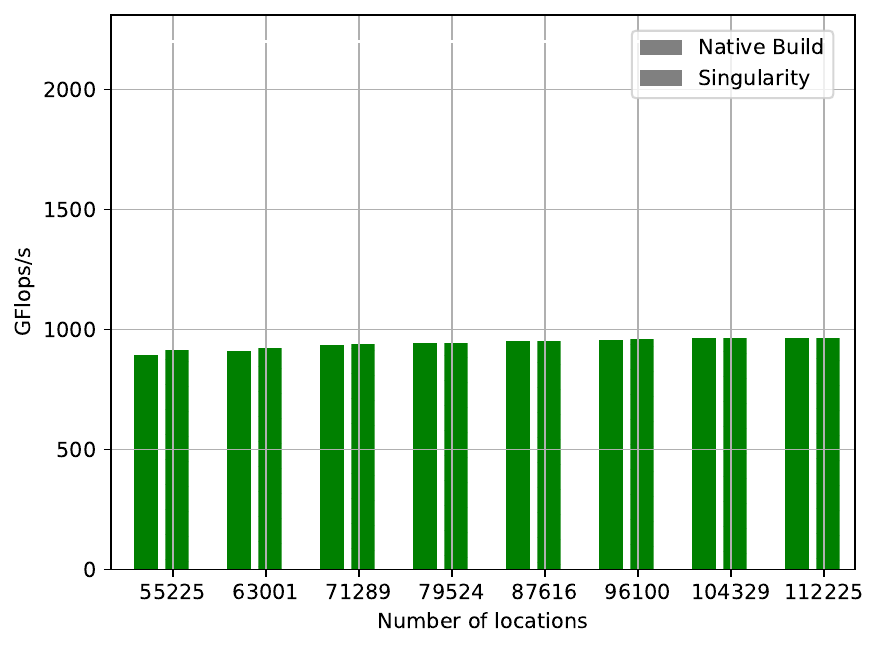}
        \hspace{2mm}
         \includegraphics[width=0.26\textwidth]{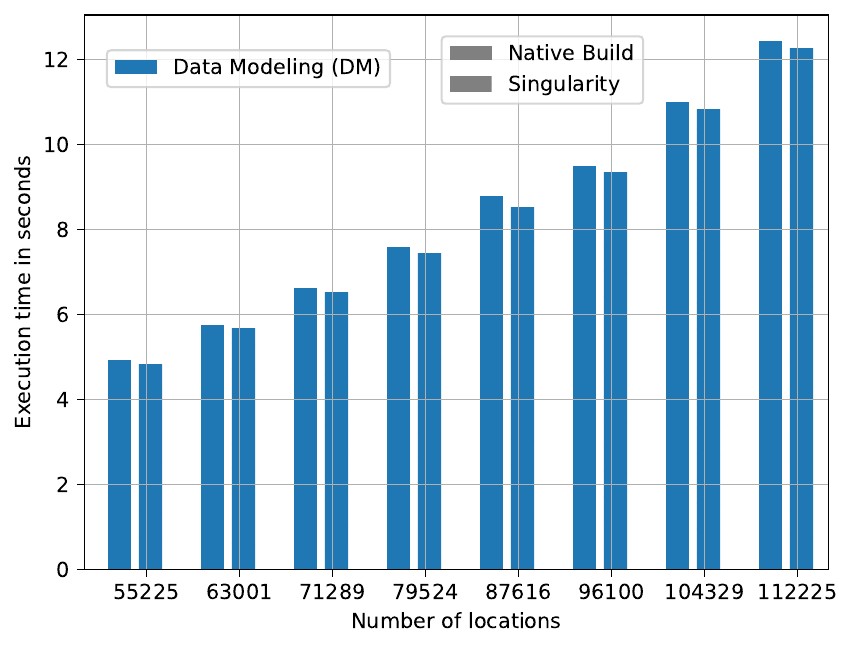}
         \caption{64-core AMD Naples.}
         \label{fig:naples}
     \end{subfigure}
        \caption{Performance assessment on CPU-based shared-memory systems.}
        \label{fig:shared-memory}
\end{figure*}

In this subsection, we aim to evaluate the performance of the Singularity images compared to the native build of the software and its dependencies using shared-memory systems. Figure~\ref{fig:shared-memory} shows the assessment results on five different CPU-based architectures. The runs include different workloads representing the number of spatial locations on the x-axis. Considering many spatial locations produces a larger covariance matrix, which requires more memory and computing resources. Each subfigure represents a specific hardware architecture and includes three subfigures. On the left, the subfigure represents the execution time to perform each operation in the software in the dense mode, where the y-axis represents the execution time in seconds. In the middle, the subfigure represents the average performance in GFlops/s (y-axis) under different workloads and in dense mode. Finally, on the right, the subfigure shows the execution time of the three operations when applying TLR approximation under various workloads.

Figure~\ref{fig:haswell} shows the performance comparison using an Intel Haswell chip. The variation in performance in the dense computation between the Singularity image and the native build is between -1\% and +0.4\%, while in TLR execution between  -1\% and +2\%. The variation when running on Intel Skylake is between -0.8\% and +1\% in the dense case, as shown in Figure~\ref{fig:skylake}. In the TLR case, the variation is between -2\% and +0.8\%. Figure~\ref{fig:icelake} shows the performance variation when the Intel Icelake chip is used. The variation between the Singularity image and the native build-in performance is between -7\% and +2\%   in the dense case and   -4\% and +2\%   in the TLR case. On the AMD Milan chip, the variation in performance is -1\% and -0.4\% between the image and native build in the dense case and between in the TLR case as shown in Figure~\ref{fig:milan}. Finally, on the AMD Naples chip, the variation in performance between -0.2\% and 1\% between the image and native build in the dense case and between -2\% and -1\% in the TLR case as shown in Figure~\ref{fig:naples}.

\begin{figure}
     \centering
        \begin{subfigure}[b]{0.5\textwidth}
         \centering
         \includegraphics[width=0.44\textwidth]{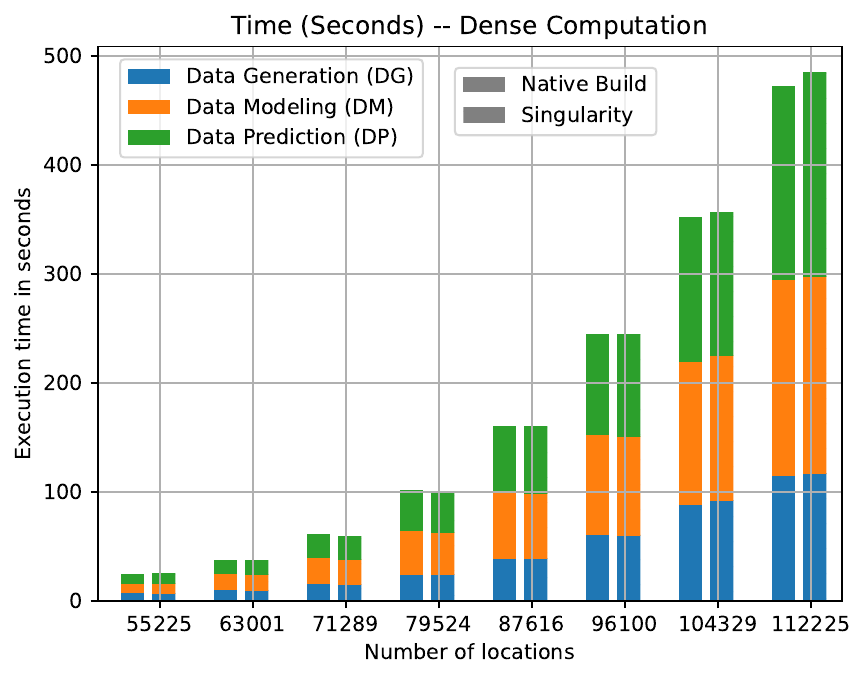}
              \hspace{4mm}
         \includegraphics[width=0.44\textwidth]{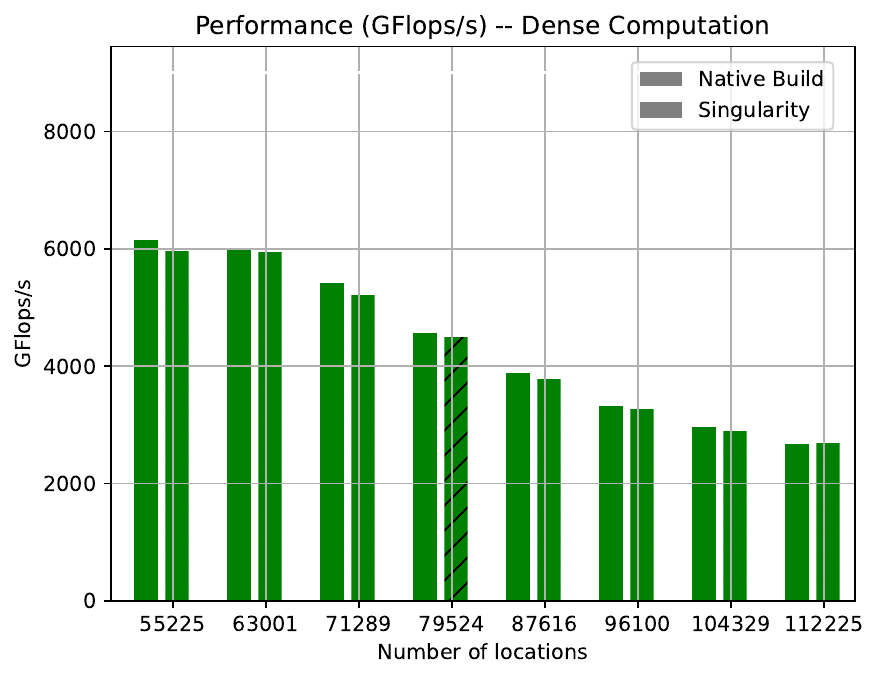}
         \caption{NVIDIA V100 + 20-core Skylake.}
         \label{fig:skylake-v100}
     \end{subfigure}
        \begin{subfigure}[b]{0.5\textwidth}
         \centering
         \includegraphics[width=0.44\textwidth]{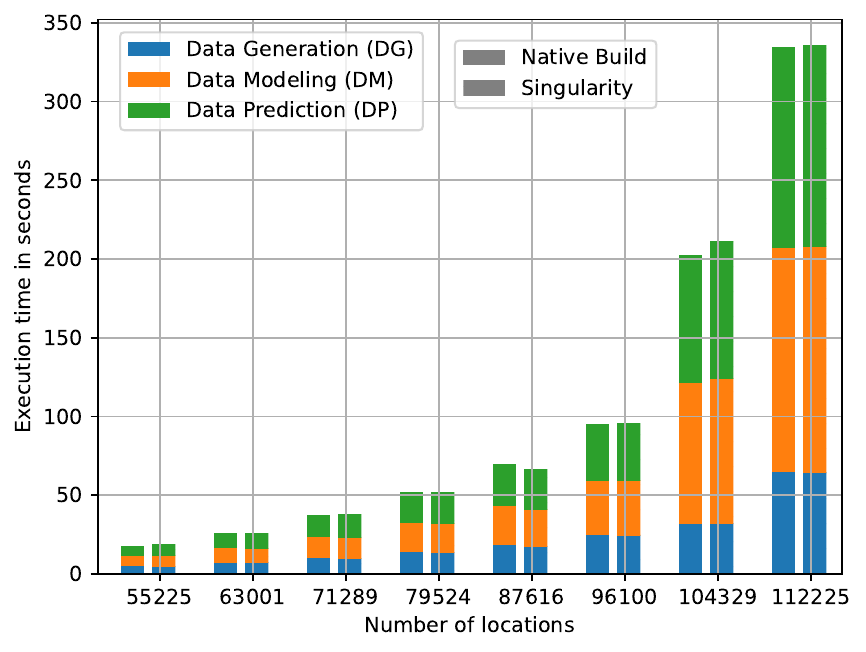}
              \hspace{4mm}
         \includegraphics[width=0.44\textwidth]{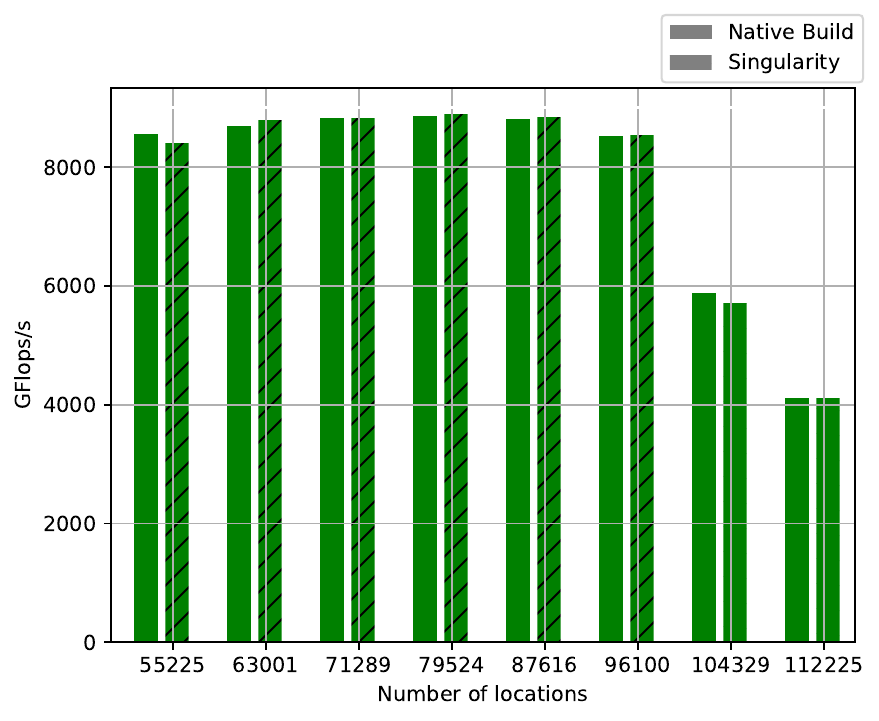}
         \caption{NVIDIA A100 + 28-core Icelake.}
         \label{fig:icelake-A100}
     \end{subfigure}
        \caption{Performance on heterogeneous shared-memory systems.}
        \label{fig:shared-memory-gpus}
\end{figure}

We also use two shared-memory systems with accelerators to assess the performance, i.e., 20-core Intel Skylake with NVIDIA V100 GPU and 20-core Intel Icelake with A100 GPU. We report the variation in performance using just the dense case since the software does not support TLR when relying on GPUs. Figure~\ref{fig:shared-memory-gpus} shows the variation in performance on the two systems. The variation in performance between the Singularity image and the native build using the NVIDIA V100 + Intel Skylake CPU is between -3\% and -1\% while using the NVIDIA A100 + Intel Icelake CPU is between -2\% and +1\% for the dense case. The degradation in performance with larger problem sizes is because of the GPU memory wall, so the software starts to use the CPU memory and move the data between the two devices when needed.

\subsection{Containers Portability and Scalability on Distributed-Memory Systems}
\begin{figure}
        \centering
        \begin{subfigure}{0.22\textwidth}
                \centering
                \includegraphics[width=\textwidth]{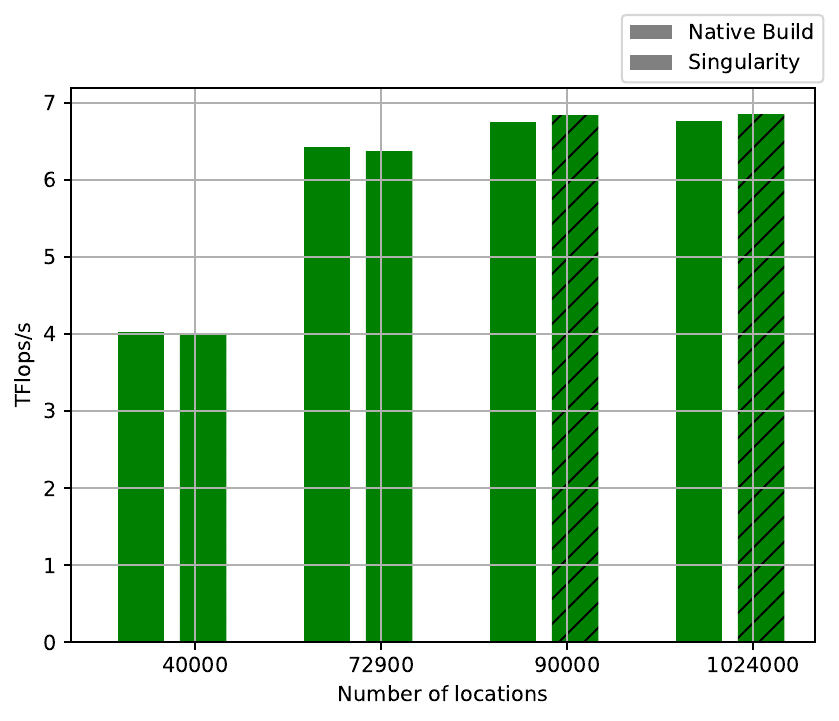}
                \caption{4 nodes.}
                \label{fig:4nodes}
        \end{subfigure}
              \hspace{4mm}
        \begin{subfigure}{0.22\textwidth}
                \centering
                \includegraphics[width=\textwidth]{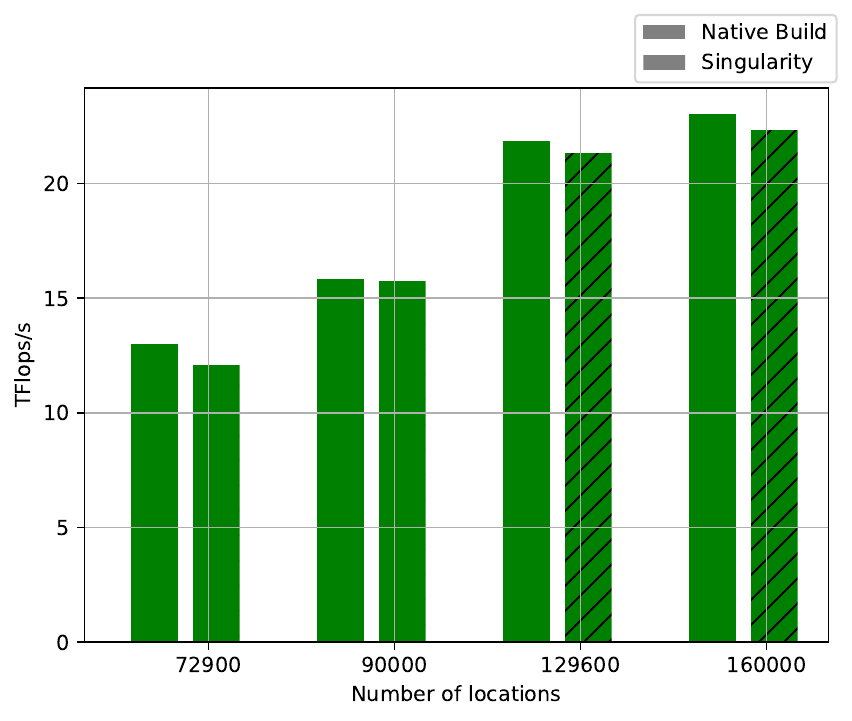}
                \caption{16 nodes.}
                \label{fig:16nodes}
        \end{subfigure}
   
        \begin{subfigure}[b]{0.22\textwidth}
                \centering
                \includegraphics[width=\textwidth]{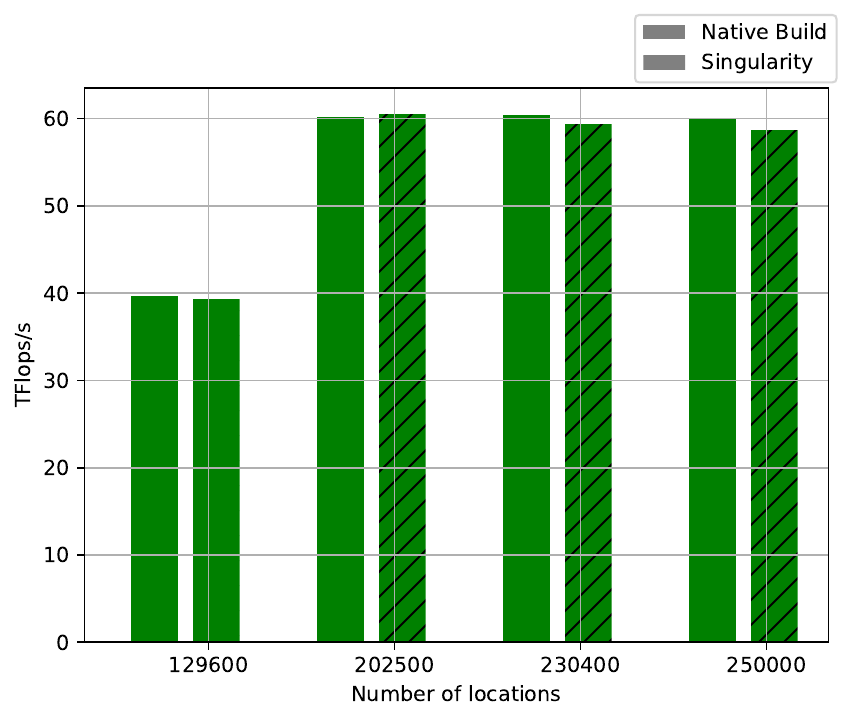}
                \caption{64 nodes.}
                \label{fig:64nodes}
        \end{subfigure}%
              \hspace{4mm}
        \begin{subfigure}[b]{0.22\textwidth}
                \centering
                \includegraphics[width=\textwidth]{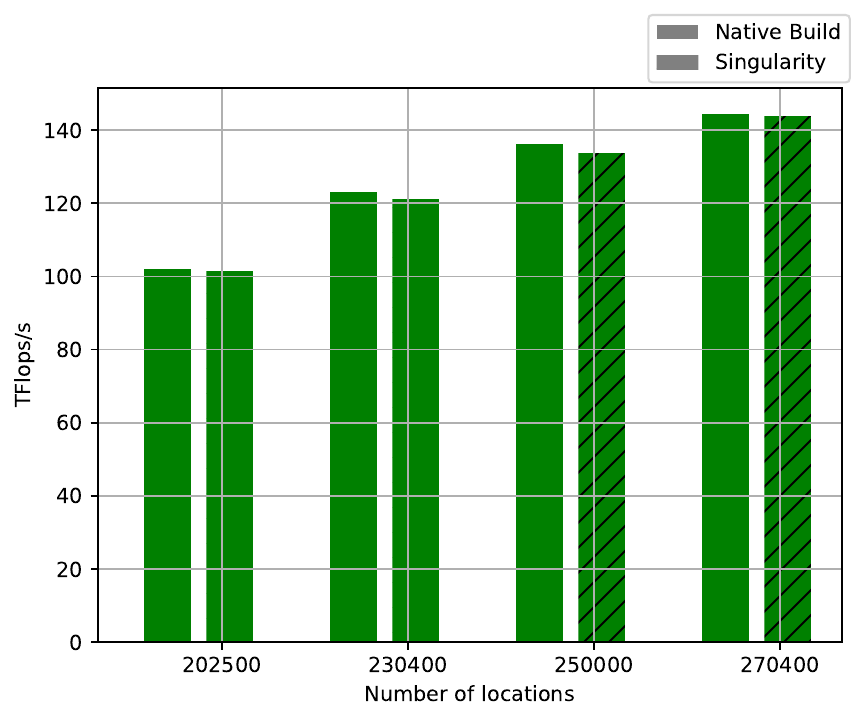}
                \caption{256 nodes.}
                \label{fig:256nodes}
        \end{subfigure}
        \caption{Performance on MareNostrum 4 supercomputer using a different number of nodes. Tile Size $ts$ has been tuned for each number of nodes. (a) $ts=760$ (b) $ts=760$ (c) $ts=760$ (d) $ts=960$.} \label{fig:dist-memory}
\end{figure}

We also conducted a set of experiments on the Marenostrum 4 supercomputer at BSC using up to $256$ nodes to validate the performance of the generated images compared to the native build. Herein, we decided to include only the dense experiments because of the page limits and because we found that the variation in performance between the generated image and the native build is the same in the TLR case. Figure~\ref{fig:dist-memory} shows the average performance in TFlops/s for the Singularity image and the native build. Figure~\ref{fig:4nodes} shows the performance in TFlops/s using $4$ nodes, which reaches close to $7$ TFlops/s. We tuned the tile size (ts), and the best performance is obtained when $ts=760$. The performance of the Singularity image varies by $\pm1$ \% compared to the native build. We can consider that there is no difference between both cases since the variation in different runs of the native build is about $\pm1$ \%. Figure~\ref{fig:16nodes} shows the performance difference on $16$ nodes. The best performance has been obtained when $ts=760$ -- up to $22$ TFlops/s. The singularity image shows $0.7$ to $7$ \% lower performance than the native build. This can also be acceptable since the variation in different runs of the native build can reach this percentage. The singularity image shows $-2$ to $+0.5$ \% compared to the native build using $ts=760$ and Using 64 nodes as shown by Figure~\ref{fig:64nodes}. The obtained performance in both cases reaches up to $60$ TFlops/s. Figure~\ref{fig:256nodes} shows that the performance of the Singularity image is $0.5$ to $2$ \% lower than the native build using $256$ nodes, and the performance can reach up to $144$ TFlops/s in both cases.

\begin{figure*}[h!]
     \centering
          \begin{subfigure}[b]{0.26\linewidth}
         \centering
         \includegraphics[width=\linewidth]{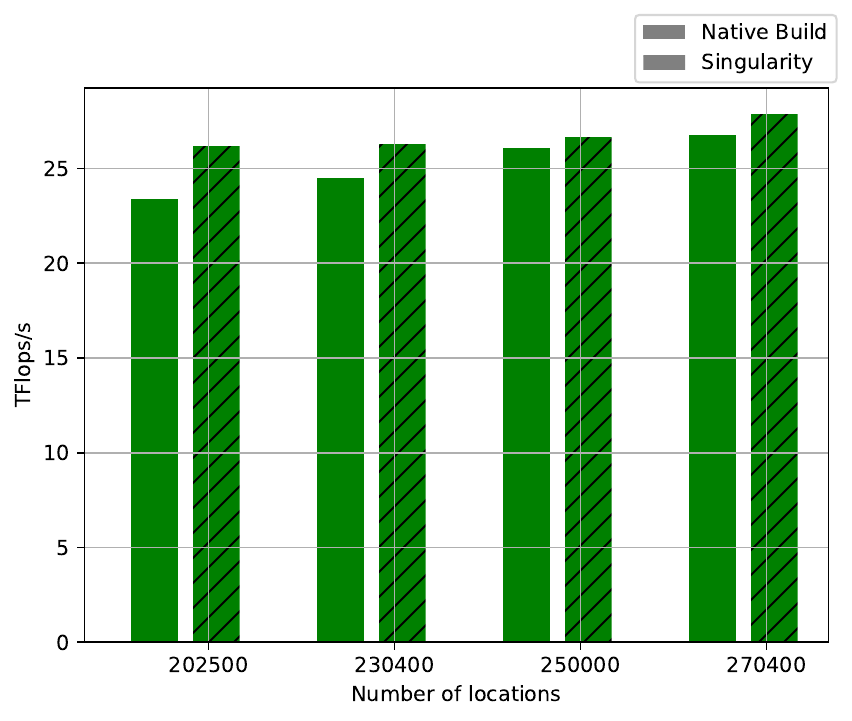}
         \caption{$ts=320$.}
         \label{fig:ts320}
     \end{subfigure}
  \hspace{6mm}
     \begin{subfigure}[b]{0.26\linewidth}
         \centering
         \includegraphics[width=\linewidth]{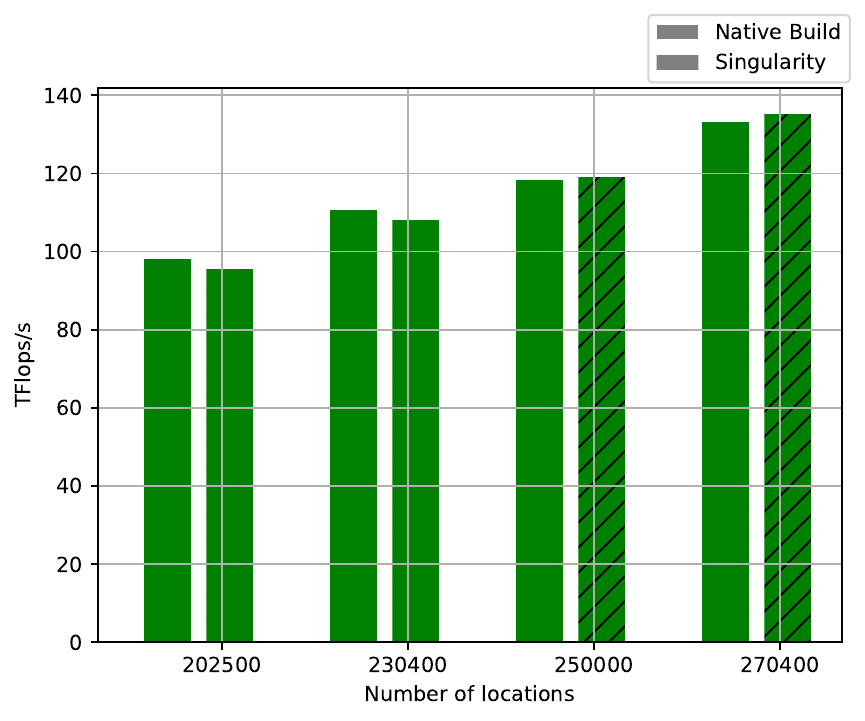}
         \caption{$ts=760$.}
         \label{fig:ts560}
     \end{subfigure}
       \hspace{6mm}
     \begin{subfigure}[b]{0.26\linewidth}
         \centering
         \includegraphics[width=\linewidth]{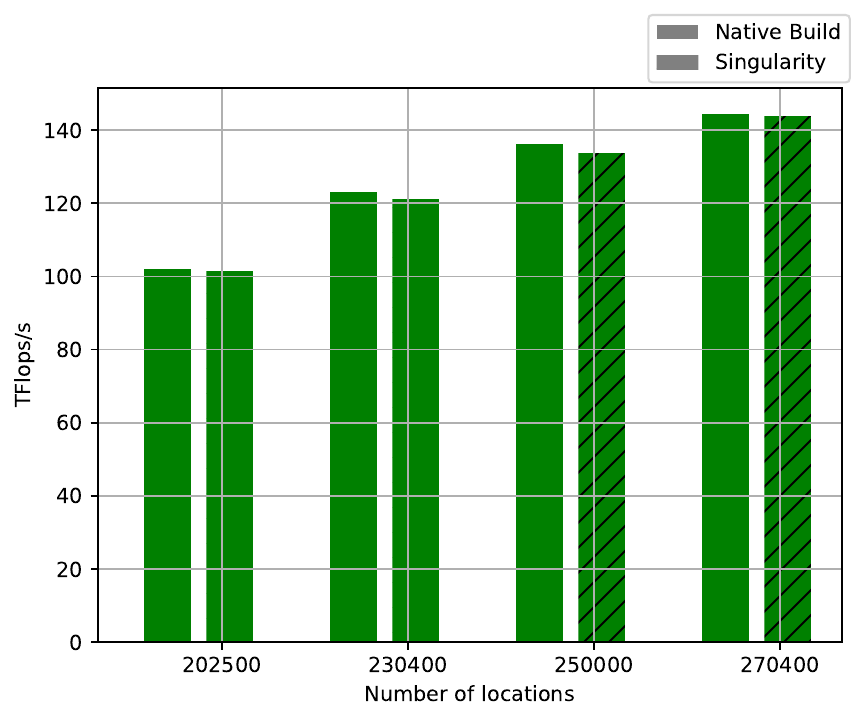}
         \caption{$ts=960$.}
         \label{fig:ts960}
     \end{subfigure}
        \caption{Performance on 256 nodes of MareNostrum 4 with different tile sizes.}
        \label{fig:shared-memory-ts}
\end{figure*}

In Figures~\ref{fig:shared-memory},~\ref{fig:shared-memory-gpus}, and~\ref{fig:dist-memory}, we tuned the tile size for each problem size in each architecture. In Figure~\ref{fig:shared-memory-ts}, we compare the performance of the HPC-ready containers with the native build using 256 nodes on  MareNostrum 4 using different tile sizes regardless of the obtained performance. The obtained results vary from around $28$ TFlops/s using $ts=320$ to about $144$ TFlops/s using $ts=960$. The variation in performance between containers and native build is between -3\% (matrix size = $202500$, and $ts=760$) to +10\% (matrix size = $202500$, and $ts=320$) in the worst case.

\section{Conclusions}
\label{sec:conclus}
Containers can be a ``gold solution'' to increase the usability of the HPC software in scientific domains through their ability to preserve performance while allowing easy portability on different HPC systems. Unlike Virtual Machines (VM), containers take up less space, consume fewer resources, and are easier to manage and scale as they can be quickly cloned and deployed. This paper assesses the portability and scalability of pre-generated Singularity images through a parallel geospatial statistics software \textit{ExaGeoStat}, which allows large-scale synthetic geospatial data generation, modeling, and prediction on manycore systems. We detail how to create HPC-ready containers based on the Spack package manager. Our results show how the pre-generated containers perform compared to the native build of the \textit{ExaGeoStat} software. In our experiments, we explore seven distinct shared-memory systems, two of which are equipped with NVIDIA accelerators. Additionally, we utilize the MareNostrum 4 supercomputer at the Barcelona Supercomputing Center to evaluate the performance of containers on a distributed-memory system. Our findings indicate that HPC-ready containers demonstrate performance levels nearly equivalent to native builds while offering enhanced portability across various architectures.

\section{Acknowledgements}
\label{sec:acks}
This work has been supported by King Abdullah University of Science and Technology (KAUST), Saudi Arabia, the Spanish Government (contract PID2019-107255GB), by the Generalitat de Catalunya (contract 2017-SGR-01414), and by the European Commission's Horizon 2020 Framework program and the European High-Performance Computing Joint Undertaking (JU) under grant agreement No 955558 and by MCIN/AEI/10.13039/501100011033 and the European Union NextGenerationEU/PRTR (project eFlows4HPC).

 \bibliographystyle{elsarticle-num} 
 \bibliography{arxiv-paper}





\end{document}